\documentclass[twocolumn,english,pre,floatfix,citeautoscript,nofootinbib]{revtex4}
\usepackage{amsbsy}
\usepackage{latexsym,epsfig,graphicx}
\usepackage{dcolumn}
\usepackage{subfigure}
\usepackage{comment}
\usepackage{color}
\usepackage[colorlinks,urlcolor=blue,citecolor=blue]{hyperref}
\usepackage{amstext}
\usepackage{amssymb}
\usepackage{setspace}
\usepackage{amsmath}
\usepackage{makeidx}
\usepackage{bm}
\usepackage{ulem}
\usepackage{multirow}
\usepackage{mathrsfs}
\usepackage{array}

\begin{document}
\title{The optical rogue wave patterns in coupled defocusing systems}
\author{Yan-Hong Qin$^{1}$}
\author{Liming Ling$^{1}$}\email{linglm@scut.edu.cn}
\author{Li-Chen Zhao$^{2,3,4}$}\email{zhaolichen3@nwu.edu.cn}
\address{$^{1}$School of Mathematics, South China University of Technology, Guangzhou 510640, China}
\address{$^{2}$School of Physics, Northwest University, Xi'an 710127, China}
\address{$^{3}$Shaanxi Key Laboratory for Theoretical Physics Frontiers, Xi'an 710127, China}
\address{$^{4}$Peng Huanwu Center for Fundamental Theory, Xi'an 710127, China}
\begin{abstract}
	
We systematically investigate rogue wave's spatial-temporal pattern in $N$ $(N\geq2)$-component coupled defocusing nonlinear Schr\"{o}dinger equations. The fundamental rogue wave solutions are given in a unified form for both focusing and defocusing cases.  We establish the quantitative correspondence between modulation instability and rogue wave patterns, which develops the previously reported inequality relation into an equation correspondence. As an example, we demonstrate phase diagrams for rogue wave patterns in a two-component coupled system, based on the complete classification of their spatial-temporal structures. The phase diagrams enable us to predict various rogue wave patterns, such as the ones with a four-petaled structure in both components. These results are meaningful for controlling the rogue wave excitations in two orthogonal polarization optical fibers.

\end{abstract}
\date{\today}

\maketitle

\section{Introduction}
Optical systems provide a good platform for studying rogue waves (RWs) \cite{ana1,Nail1}, which could endanger marine navigation and optical communications. Some rational solutions of $N$-component nonlinear Schr\"{o}dinger equations (NLSE) have been widely used to describe RW phenomena in nonlinear optics fibers with taking no higher-order effects \cite{exp0,exp1}, and other nonlinear systems \cite{ana4,exp2,exp3,ana3,Zhenya,Rev}. In most previous studies, RW solutions were obtained in focusing NLSE  \cite{Nail2,He1,ling,Baronio1,lingrw1,zhaorw1,zhaorw2,zhaorw3,Zhenya,Liu1}. The fundamental RW for a scalar NLSE ($N=1$) is always eye-shaped (ES). However, vector RWs ($N\geq 2$) can involve an eye-shaped one \cite{Baronio1,lingrw1}, an anti-eye-shaped one (AES) \cite{zhaorw1}, and a four-petaled one (FP) \cite{zhaorw2}, since vector systems allow for energy transfer between the coupled waves. The nonlinear superposition of these fundamental RWs can produce more diverse structures, which refer to the higher-order RWs or multiple RWs \cite{He2,ling2,ling3,Liu2,Gao}. So far, modulation instability (MI) is believed to play an important role in RWs' excitations \cite{MI,MI1,MI2,zhaoMI}. Moreover, the equation correspondence between the fundamental RW solution and the dispersion relation of MI was established in the focusing $N$-component NLSE \cite{lingMI}, which can be used to interpret the RW patterns perfectly. It predicts that there are mainly three different pattern types for arbitrary N-component coupled focusing NLSE in integrable cases.

On the contrary, RW can not exist in the scalar defocusing NLSE, since there is no MI on the plane wave background. A few pieces of literature recently reported that vector RWs could exist in integrable two-component defocusing systems \cite{defocusing1,defocusing2,defocusing3,defocusing6,chen1}, with some certain constraints on the two plane wave backgrounds. Those theoretical results have motivated experiments to observe AES-AES (i.e., dark-dark) RW in nonlinear optical fibers \cite{defocusing4,defocusing5}. The AES-AES RW refers to the two components admitting the AES RW pattern. However, the classification of vector RW patterns is still unclear, let alone the $N$ $(N>2)$-component case. On the other hand, the inequality relation between the RW solutions and MI was suggested \cite{defocusing1}, which inspired many discussions for the generation mechanism of RWs \cite{MI3,zhaoMI,lingMI}. It is very essential to establish the equation correspondence between MI and RWs' spatial-temporal structures for defocusing cases, which is also meaningful for controllably exciting various vector RWs in experiments.

In this work, we systematically study the patterns of vector RWs in coupled defocusing nonlinear systems, based on the general exact RW solutions of $N (N \geq2)$-component defocusing NLSE. Importantly, we successfully establish the equation correspondence between MI and RW solutions, which provides the quantitative mechanism of vector RWs in defocusing regime. For example, we present phase diagrams to illustrate the families of two-component RWs in defocusing and focusing regimes. FP-FP patterns are predicted based on the phase diagrams, in contrast to the previously observed AES-AES RW \cite{defocusing4,defocusing5}. We further discuss the possibility to experimentally observe them from weakly localized perturbations in two orthogonal polarization optical fibers.

The paper is organized as follows. In Sec.~\ref{sec2}, we first present the unified RW solutions in an arbitrarily $N$ $(N\geq2)$-component coupled NLSE in both defocusing and focusing regimes. Then, the systematic classification of vector RWs in a two-component case has been demonstrated clearly, by giving three sets of phase diagrams for RW patterns. In Sec.~\ref{sec3}, we obtain the equation correspondence between the dispersion relation of MI and the existing condition of RW patterns, which implies that the resonance perturbations in MI regions can be used to generate RW controllably and directly. In Sec.~\ref{sec4}, we discuss the possibilities of observing vector RW patterns in defocusing nonlinear fibers. For example, the numerical simulations demonstrate that FP-FP RW can be excited perfectly by combining the phase diagrams and using the resonance perturbations. Finally, we summarize our results in Sec.~\ref{sec5}.

\section{Rogue wave patterns in coupled nonlinear Schr\"{o}dinger systems}\label{sec2}

\subsection{The unified vector rogue wave solutions}

We start with $N$-mode coupled nonlinear fibers, described by the following dimensionless integrable $N$-component NLSE:
\begin{eqnarray}
\label{model}
\mathrm{i}\mathbf{\Phi}_\xi+\frac{1}{2}\mathbf{\Phi}_{\tau\tau}+\sigma\mathbf{\Phi}\mathbf{\Phi}^{\dag}\mathbf{\Phi}=0,
\end{eqnarray}
where $\mathbf{\Phi}=(\phi_1,\phi_2,...,\phi_N)^{\mathrm{T}}$. For the vector cases, $N\geq2$ ($N$ is an integer) is inevitable. The symbols ${\mathrm{T}}$ and ${\dag}$ represent transpose and Hermite conjugation of a matrix, respectively. $\phi_i(\tau,\xi)$ represents the complex slowly varying envelope of the $i$-th mode. Here $\xi$ and $\tau$ denote the propagation distance and retarded time. It should be pointed out that the meanings of these two variables depend on the particular applicative context. $\sigma=1$ and $\sigma=-1$ correspond to the focusing and defocusing nonlinearity, respectively.

\begin{figure}[ht!]
	\centering
	{\includegraphics[width=80mm]{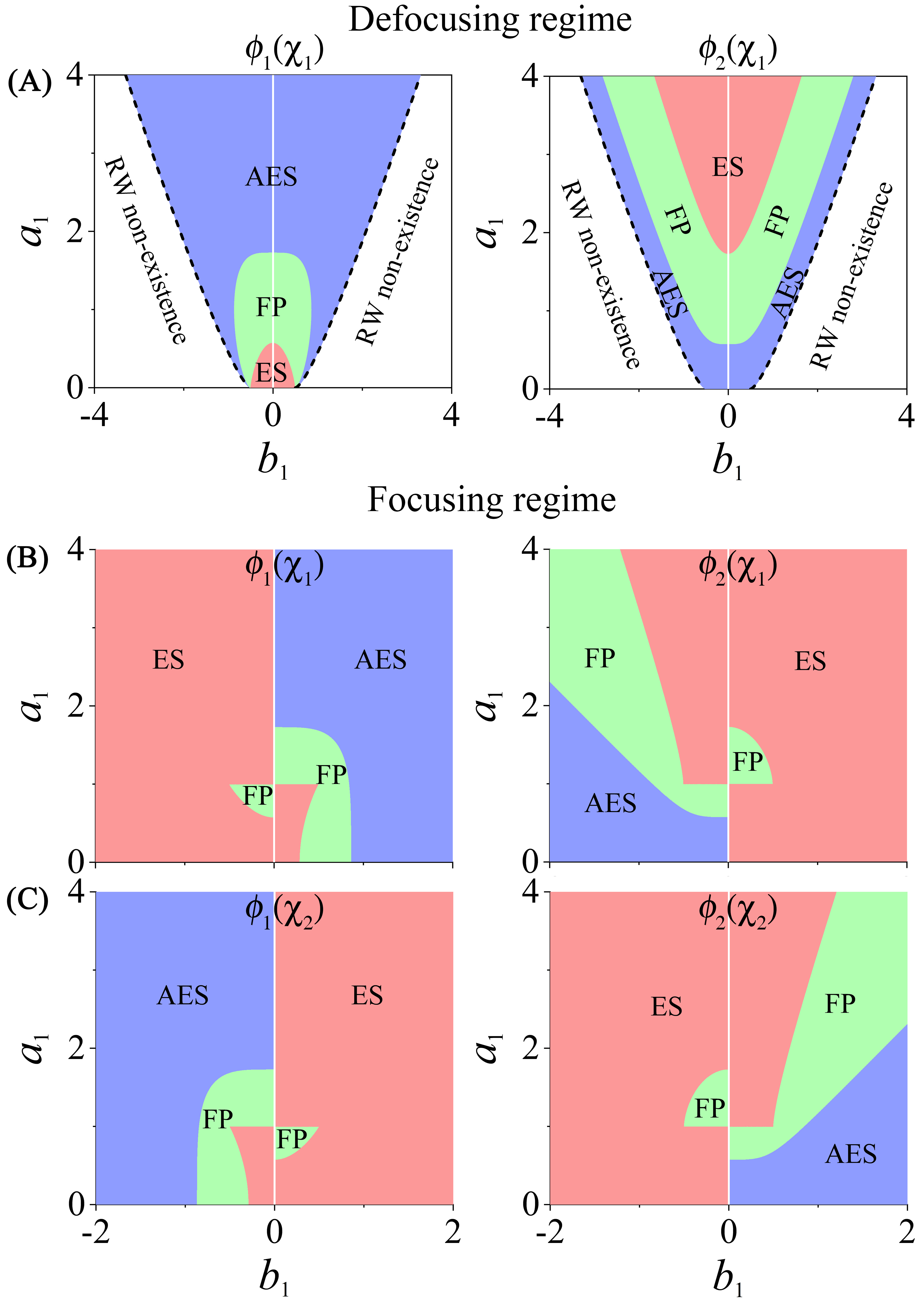}}
	\caption{The phase diagrams of two-component vector RW patterns in $(b_1,a_1)$ plane. (A): Defocusing case. (B) and (C): Two sets of phase diagrams in the focusing regime. The white regions and white lines ($b_1=0$) denote RW non-existence. Black dashed curves depict the critical conditions of RW excitations. The pink, green, and blue regions denote the RW solutions with ES, FP, and AES patterns. The parameters are $a_2=1,b_2=-b_1$.}\label{Fig1}
\end{figure}

By conducting Darboux transformation method \cite{ling2,ling3}, we demonstrate that the fundamental vector RW solutions of Eq.~\eqref{model} with focusing or defocusing nonlinearity have a unified form
\begin{equation}\label{RW}
\phi_i=a_i\Big\{1+\frac{2\mathrm{i}(\chi_R+b_i)(\tau+\chi_R\xi)-2\mathrm{i}\chi_I^2\xi-1}
{\varpi_i\left[(\tau+\chi_R\xi)^2+\chi_I^2\xi^2+1/(4\chi_I^2)\right]}\Big\}\mathrm{e}^{\mathrm{i}\theta_i},
\end{equation}
where $\varpi_i=(\chi_R+b_i)^2+\chi_I^2$, $\theta_i=b_i\tau+\big(\sigma\sum_{i=1}^{N}a_i^2-\frac{b_i^2}{2}\big)\xi$, $i=1,2,\cdots,N$. The parameters $a_i$ and $b_i$ are the amplitude and frequency of the background in the $i$-th component, respectively. $\chi_R=\Re(\chi)$, $\chi_I=\Im(\chi)$, and $\chi$ is a complex root of the following algebraic equation:
\begin{eqnarray}\label{branch}
	1+\sigma\sum\limits_{i=1}\limits^{N}\frac{a_i^2}{(\chi+b_i)^2}=0,
\end{eqnarray}
which is the determining equation for RW solutions. For the focusing case ($\sigma=1$), the above algebraic equation is positive definite and always admits $N$ pairs of complex conjugate roots \cite{lingMI}. On the contrary, this polynomial is non-positive definite in the defocusing regime ($\sigma=-1$), which confirms the existence of at least two real roots, namely, the Eq.~\eqref{branch} admits $N-1$ pairs of complex conjugate roots at most. Naturally, RW solutions cannot be allowed in scalar-defocusing cases. Therefore, to obtain RW solutions in defocusing case, the amplitudes and frequencies of vector background must be strictly and adequately designed to get a complex root of $\chi$, instead of being free parameters in the focusing cases.

Some RW solutions have been presented in two-component defocusing nonlinear systems \cite{defocusing1,defocusing2,chen1,defocusing3,defocusing6}, such as AES-ES (dark-bright) RW and AES-AES RW. However, in the defocusing case, the classification of RW patterns has yet to be thoroughly investigated, and the parameter domains for different vector RWs have not been analyzed. The fundamental RW patterns in the focusing case had been studied clearly, mainly including the ES, AES, and FP ones \cite{lingMI}. The difference and relation between the vector RWs in focusing and defocusing cases still need to be discussed.  Therefore, we further systematically investigate the pattern classification of vector RWs in a two-component system.

\subsection{The pattern classification of fundamental vector rogue waves}

In the defocusing regime with $N=2$, we get the discriminant to the existence of RW is
$\big[a_1^2+a_2^2-(b_1-b_2)^2\big]^3+27 a_1^2a_2^2(b_1-b_2)^2>0$, which ensures the presence of a pair of complex conjugated roots of the Eq.~\eqref{branch}. Without loss of generality, we present the phase diagrams of vector RW solutions in Fig.~\ref{Fig1}(A) with the setting $a_2=1$ and $b_2=-b_1$. Here, we denote the paired roots satisfying the discriminant as $\chi_{\mathrm{1}}$. The RW can not be excited in the parameter space marked by the white regions and white lines in the $(b_1,a_1)$ plane. The critical condition depicted by black dashed curves is calculated as $a_1=[3(4b_1^2)^{\frac{1}{3}}-3(4b_1^2)^{\frac{2}{3}}+4b_1^2-1]^{1/2}$. By analyzing the extreme points of RW solutions, we get that the solutions \eqref{RW} with $\frac{(\chi_R+b_i)^2}{\chi_I^2}\leq\frac{1}{3}$, $\frac{1}{3}<\frac{(\chi_R+b_i)^2}{\chi_I^2}<3$ and $\frac{(\chi_R+b_i)^2}{\chi_I^2}\geq3$ can form the ES RW (pink areas), FP RW (green areas), and AES RW (blue areas), respectively. The phase diagram Fig.~\ref{Fig1}(A) demonstrates that the family of vector RWs in the defocusing cases includes AES-ES RW, AES-FP RW, AES-AES RW, and FP-FP RW. Here, AES-ES RW and ES-AES are regarded as the same vector RW; the definitions of other vector RWs are similar to this case. In Ref.~\cite{defocusing1}, only the AES-AES RW and AES-ES RW were obtained in the defocusing regime. Figure \ref{Fig1}(A) also indicates that ES-ES RW and ES-FP RW are not allowed in the defocusing regime. It is seen that the probability of forming AES far outweighs that of ES RW and FP RW in the component $\phi_1$. When the amplitude $a_1>1.73$, the component $\phi_1$ only allows the AES RW. On the contrary, the probability of exciting ES RW and FP RW in the component $\phi_2$ is much greater than that of AES RW. Remarkably, the FP-FP RW can be produced in a small parameter domain in regions with amplitude $a_1\in[0.58,1.73]$. This phase diagram clearly demonstrates the existence conditions of vector RW solutions and the classification of RW patterns in defocusing two-component systems, which has never been determined in previous works. Similar phase diagrams can be obtained by taking other parameters.

On the contrary, in the focusing regime, the excitations of vector RWs have no restriction on the amplitudes and frequencies for the vector background. There are always two pairs of complex conjugate roots (denoted as $\chi_{\mathrm{1}}$ and $\chi_{\mathrm{2}}$) of Eq.~\eqref{branch}, which give rise to two different sets of phase diagrams, as shown in Fig.~\ref{Fig2}(B) and (C). As it can be seen, the RW patterns can be always excited in $(b_1,a_1)$ plane, in sharp contrast to that of in defocusing case (compared to Fig.~\ref{Fig1}(A)). In this case, the vector RWs include ES-ES RW, ES-FP RW, ES-AES RW, and FP-FP RW. Strikingly, the FP-FP RW only can be generated in a small parameter space characterized by the two small green fan-shaped areas. Furthermore, in most cases, the types of vector RWs in the case of $\chi_1$ is different from that of $\chi_2$ for the same values of $a_1$ and $b_1$, rather than the exchange of RW types between the two components. For example, there are AES-ES RW for $\chi_1$ and ES-FP RW for $\chi_2$ with choosing $b_1=1.5, a_1=3$; FP-FP RW for $\chi_1$ and ES-ES RW for $\chi_2$ with setting $b_1=0.25,a_1=1.4$. Therefore, in the focusing regime, two different RW types can be obtained in the same vector background, which is not admitted in defocusing case. In Ref.~\cite{zhaoEPL}, only a phase diagram similar to Fig.~\ref{Fig2}(C) was presented, where the parameter regions for FP-FP RW had been lost.

\section{The equation correspondence between modulation instability and rogue wave patterns}\label{sec3}

Previous studies suggested that baseband MI can be seen as the origin for RW formations in defocusing nonlinear regime \cite{defocusing1,MI3}. However, the correspondence relation between the RW existence condition and baseband MI was an inequality form, which was also obtained in the two-component case. In the defocusing case, the quantitative correspondence between the MI and the existing condition of RW solutions is still lacking. The equation correspondence for the focusing N-component NLSE \cite{lingMI} motivates us to look further for similar results for defocusing cases.

We revisit the standard MI analysis. The linearized stability of perturbations on the plane wave solution can be obtained by adding weak perturbations with Fourier modes. Then, a perturbed vector background is written as $\phi_i=\phi_i^{[0]}[1+p_i(\tau,\xi)]$ ($i=1,2,\cdots,N$). Here $\phi_i^{[0]}=a_i \mathrm{exp}(\mathrm{i}\theta_i)$ are vector background solutions, and $p_i(\tau,\xi)$ are small perturbations which satisfy the linear equation $\mathrm{i}(p_{i,\xi}+b_ip_{i,\tau})+\frac{1}{2}p_{i,\tau\tau}+\sigma\sum_{l=1}^{N}[a_l^2(p_l+p_l^*)]=0$. The asterisk means complex conjugate. We suppose the perturbations $p_i(\tau,\xi)$ have the form
$p_i(\tau,\xi)=p^{*}_{i,-k}\mathrm{exp}[-\mathrm{i}\mu_k(\tau+\Omega_k^{*}\xi)]+p_{i,k}\mathrm{exp}[\mathrm{i}\mu_k(\tau+\Omega_k\xi)]$. Then, we get $\mathcal{K}\mathcal{P}=0$, where $\mathcal{K}=\mathrm{diag}[(-\Omega_k\!-\!b_1\!-\!\frac{1}{2}\mu_k)\mu_k,(\Omega_k\!+\!b_1\!-\!\frac{1}{2}\mu_k)\mu_k,\cdots,(-\Omega_k\!-\!b_N\!-\!
\frac{1}{2}\mu_k)\mu_k,(\Omega_k\!+\!b_N\!-\!\frac{1}{2}\mu_k)\mu_k]\!-\!\mathcal{I}\mathcal{A}$ and   $\mathcal{P}=(p_{1,k},p_{1,-k},\cdots,p_{N,k},p_{N,-k})^{\mathrm{T}}$, with $\mathcal{I}=(1,1,\cdots,1,1)^{\mathrm{T}}$, and  $\mathcal{A}=(\sigma a_1^2,\sigma a_1^2,\cdots, \sigma a_N^2, \sigma a_N^2)$. The determinant of matrix $\mathcal{K}$ is
$\mathrm{det}(\mathcal{K})=\mu_k^{2N}\prod_{l=1}^{N}\big[\frac{1}{4}\mu_k^2-(\Omega_k+b_l)^2\big]\times\big[1+\sigma \sum_{l=1}^{N}\frac{a_l^2}{(\Omega_k+b_l)^2-\frac{1}{4}\mu_k^2}\big]$. To get the nonzero solution of vector $\mathcal{P}$, the determinant $\mathrm{det}(\mathcal{K})$ must be equal to zero, which is the dispersion relation for linearized disturbance, i.e.,
$1+\sigma \sum_{l=1}^{N}\frac{a_l^2}{(\Omega_k+b_l)^2-\frac{1}{4}\mu_k^2}=0$. The roots $\Omega_k$ with a nonzero imaginary part correspond to linearly unstable modes, with growth rate $|\mathrm{Im}(\mu_k\Omega_k)|$. It was suggested that RW came from the resonance perturbations in MI regions \cite{zhaoMI}, which refers to that both the dominant frequency and propagation constant of perturbation are equal to those of the background. Inspired by this, we take the limit $\mu_k\rightarrow0$ to address the dispersion relation of MI in the defocusing case, which can be expressed as
\begin{eqnarray}
	\label{MI2}
	1+\sigma \sum\limits_{l=1}\limits^{N}\frac{a_l^2}{(\Omega_k+b_l)^2}=0.
\end{eqnarray}
Surprisingly, this MI dispersion relation form is consistent with the determining equation Eq.~\eqref{branch} of RW solutions, which means the equation correspondence between the two ($\chi=\Omega_k$). $\Re(\Omega_k)$ stands for the evolution energy of the perturbation, and $\Im(\Omega_k)$ denotes the growth rate of a perturbation responsible for the formation of RW. The above dispersion relation with $\sigma=-1$ is non-positive definite, which leads to two types of dispersion relations. One is $\Im(\Omega_k)\equiv0$, which stands for the linearly stable mode (denoted as modulational stability (MS) branch). But the other can admit $\Im(\Omega_k)\neq0$ under some special constraints on the amplitudes and frequencies of the background, which corresponds to linearly unstable mode (denoted as MI branch). Therefore, only if the weak perturbations are set in MI branch, the RW can be excited, as the phase diagram shown in Fig.~\ref{Fig1}(A). Then, based on the equation correspondence between the MI and RW solution, i.e., $\chi=\Omega_k$, the RW patterns evolved from the MI branch can be predicted. However, when the weak perturbations choose MS branch, the dynamical evolutions could produce the vector dark solitons on the plane wave background \cite{Biondini1,lingdark}. For example, when $N=1$ with $\sigma=-1$, it is easy to get $\Omega_k=\pm a_1-b_1$. Therefore, only the MS branch exists in scalar defocusing nonlinear systems, which admit dark solitons rather than RW \cite{lingdark,Biondini2}. For two-component defocusing systems, both the MS and MI branches can exist, so dark soliton can coexist with RW or breather \cite{defocusing3}.

For the focusing case, i.e., $\sigma=1$, the above dispersion relation Eq.~\eqref{MI2} is positive definite, which leads to the roots $\Omega_k$ with a nonzero imaginary part. Therefore, there are always $N$ MI branches in the $N$-component case. By choosing any one of the MI branches, the RW patterns can be generated from the localized perturbation in the background. Meanwhile, different RW patterns can be excited on the same background for vector cases, since the existence of multiple MI branches, as the example shown in Fig.~\ref{Fig1}(B) and (C). The $N$-component focusing system can possess $N$ different patterns in each component at most. On the contrary, in the defocusing case, each component can admit $N-1$ different RW patterns at most in an $N$-component system.

Eq.~\eqref{branch} and Eq.~\eqref{MI2} have confirmed that the quantitative correspondence between the MI and RW solution exists in both the defocusing and focusing nonlinear systems, which develops the previously reported inequality relation \cite{defocusing1} into equation correspondence. It indicates that the RW patterns can be excited conveniently and controllably by using the weak resonance perturbations in MI regions.

\section{The possibilities of observing vector rogue wave patterns in defocusing nonlinear fibers}\label{sec4}

\begin{figure}[htp!]
\centering
{\includegraphics[width=85mm]{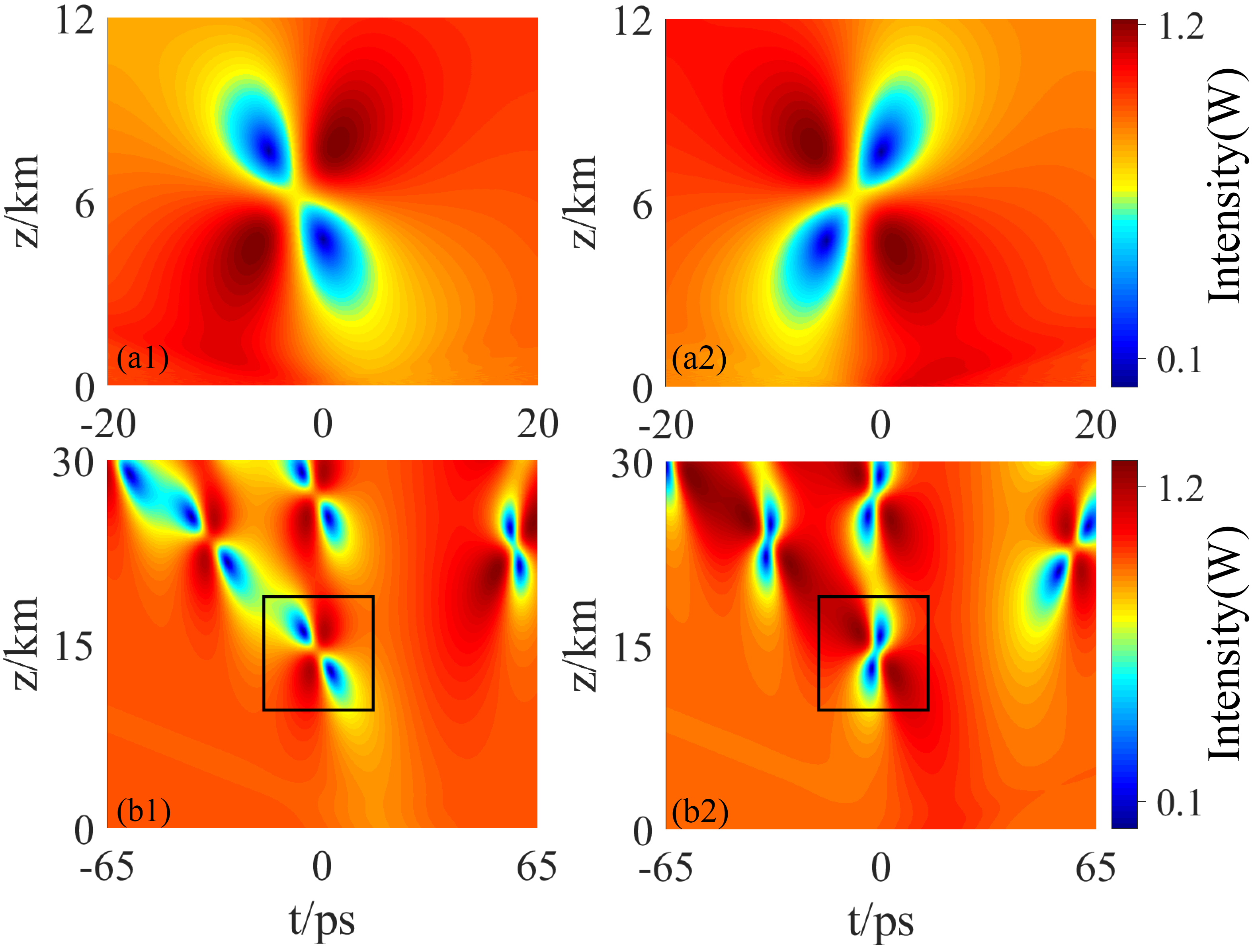}}
\caption{Contour plots of optical intensity in two orthogonal polarization waves ($|\Phi_1(t,z)|$ (left) and $|\Phi_2(t,z)|$ (right)), describing the numerical excitation of optical FP-FP RW. (a1-a2): The initial condition given by an exact one with Gaussian white noises and parameter deviation of $\chi$. (b1-b2): Numerical evolution from a plane wave background perturbed by a weak Gaussian pulse perturbation. The parameters of the vector background are $a_1=1,b_1=-b_2=0.4$.}\label{Fig2}
\end{figure}

The above theoretical results can be used to excite vector RWs in a controllable way. The phase diagram Fig.~\ref{Fig1}(A) guides one to design the amplitudes and frequencies of the two plane wave backgrounds for generating a certain RW pattern. The equation correspondence between MI and RW indicates that the resonance perturbations in the MI regions can be used to generate RW conveniently. Namely, we can excite RW by adding an arbitrary localized perturbation instead of approaching the ideal initial condition given by the exact solution.

Recently, the optical dark-dark RWs have been observed experimentally \cite{defocusing4,defocusing5} in two orthogonally polarized optical fibers \cite{PMI}. We set the parameter settings to be identical to Ref.~\cite{defocusing5} to discuss the possibilities of experimentally exciting other types of vector RWs in two-component defocusing case. The propagation of two orthogonally polarized optical pump waves at a relative frequency offset $\Delta$ in a random weakly birefringent telecom fiber is described by the dimensional Manakov equations, $\mathrm{i}\Phi_{1,z}+\mathrm{i}\delta/2\Phi_{1,t}-\beta_2/2\Phi_{1, tt}+\gamma(|\Phi_1|^2+|\Phi_2|^2)\Phi_1=0$ and $\mathrm{i}\Phi_{2,z}-\mathrm{i}\delta/2\Phi_{2,t}-\beta_2/2\Phi_{2, tt}+\gamma(|\Phi_1|^2+|\Phi_2|^2)\Phi_2=0$. Here $z$ and $t$ denote the dimensional propagation distance and retarded time coordinates, respectively. $\Phi_1$ ($\Phi_2$) is a slow (fast) wave. $\delta$=$\beta_2\Delta$ is associated with their group-velocity mismatch owing to normal group-velocity dispersion. $\beta_2$ and $\gamma$ are the group-velocity dispersion and the effective Kerr nonlinear coefficient, respectively. There is a transformation between the dimensional and dimensionless models, i.e., $\Phi_1(t,z)=\sqrt{P_0}\phi_1\exp{(\mathrm{i}\frac{\delta}{2\beta2}t-\mathrm{i}\frac{\delta^2}{8\beta2}z)}$, $\Phi_2(t,z)=\sqrt{P_0}\phi_2\exp{(-\mathrm{i}\frac{\delta}{2\beta2}t-\mathrm{i}\frac{\delta^2}{8\beta2}z)}$, $t=\tau t_0$, $z=-\xi z_0$, with $z_0=(\gamma P_0)^{-1}$, and $t_0=\sqrt{\beta_2z_0}$.  The parameters of this system are set as $\beta_2=18\mathrm{ps}^2/\mathrm{km}, \gamma=2.4\mathrm{W}^{-1}\mathrm{km}^{-1}, \Delta=100\mathrm{GHz}$, and $P_0=\frac{9\pi^2\Delta^2\beta_2}{8\gamma}10^{-6}$. Then, the characteristic length is $z_0\approx0.5\mathrm{km}$, and the timescale is $t_0\approx3\mathrm{ps}$.

The phase diagram Fig.~\ref{Fig1}(A) shows that the FP-FP RW is mainly admitted in the regions with amplitudes $a_1\in[0.58,1.73]$, $a_2=1$, and frequencies $b_1\in[-0.76,0.76]$, $b_2=-b_1$ for the two plane wave backgrounds. We choose, for example, the background parameters $a_1=a_2=1,b_1=-b_2=0.4$ to create FP-FP RW expectantly. We first consider the numerical evolutions from the initial condition given by the exact solution Eq.~\eqref{RW} with the parameter deviation and adding Gaussian white noises. We set $\Phi_{j,noise}(t,0)=\Phi_{j}(t,-9;\chi)[1+\mathrm{rand}(1)/10\times\exp(-t^2/4)]$ $(j=1,2)$ and $\chi=\chi_1+\mathrm{rand}(1)/20$, where $\mathrm{rand}(1)$ is random complex numbers with norm less than 1. $\chi_1=\mathrm{i}0.347311$ is determined by the Eq.~\eqref{branch}. The simulation results shown in Fig.~\ref{Fig2}(a1-a2) present the typical FP-FP RW patterns. However, it is inconvenient to prepare such initial states in real experiments.

Importantly, the equation correspondence between MI and RW indicates that the resonance perturbations in the MI regions can be used to generate RW conveniently. This means that we can excite RW by adding an arbitrary localized perturbation, instead of approaching the ideal initial condition given by the exact solution. Therefore, for example, we conduct numerical simulations to generate FP-FP patterns, by adding a weak Gaussian pulse perturbation on the vector background \cite{zhaoMI,Gao}. Namely,  $\Phi_j(t,0)=a_j\exp(\mathrm{i}\theta_j)\sqrt{P_0}\big\{1+\varepsilon_j\exp[-\frac{(t-\delta_j)^2}{w_j^2}]\big\}$ $(j=1,2)$, where $\varepsilon_j$, $\delta_j$ and $w_j$ denote the amplitude, offset and width of perturbation in $i$-th mode, respectively. We illustrate the numerical evolutions in Fig.~\ref{Fig2}(b1-b2) with choosing $\varepsilon_1=-\varepsilon_2=-0.1,w_1=w_2=5,\delta_1=\delta_2=6$, and the same vector background with Fig.~\ref{Fig2}(a1-a2). As the phase diagrams predicted above, the FP-FP RW successfully emerges at the propagation distance $z\approx14\mathrm{km}$ (enclosed by a black box), in contrast to the previously observed dark-dark RW \cite{defocusing4,defocusing5}. It suggests that there are many possibilities to observe various vector RWs in real experiments, by combining the phase diagrams and the resonance MI theory.

\section{Conclusion}\label{sec5}

In summary, the patterns of optical vector RWs in the defocusing regime have been systematically classified. We establish the equation correspondence between MI and exact RW solutions, which provides an important supplement for the results in the focusing regime \cite{zhaoMI,lingMI}. We present the phase diagrams to illustrate the families of vector RW patterns in a two-component case, which can guide one to design appropriate resonance perturbations to observe them controllably in actual experiments. For example, we numerically demonstrate that the FP-FP patterns can be excited from a weak localized perturbation in two orthogonal polarization optical fibers. The results could be used to realize controllable RW pattern excitations in nonlinear coupled systems.

\textit{Note added}. Recently, the authors became aware that the one-to-one correspondence between the baseband MI and the RW solutions was suggested in a different model \cite{MIchen}.

\section*{ACKNOWLEDGMENTS}
Y.-H. Qin was supported by the China Postdoctoral Science Foundation (Contract No. 2021M701255), and the National Natural Science Foundation of China (Contract No. 12147170). Liming Ling is supported by the National Natural Science Foundation of China (Grant No. 12122105). L.-C. Zhao was supported by the National Natural Science Foundation of China (Contracts No. 12022513 and No. 11775176), and the Major Basic Research Program of Natural Science of Shaanxi Province (Grant No. 2018KJXX-094).


\begin{thebibliography}{99}
	\bibitem{ana1} D. H. Peregrine, Water waves, nonlinear Schr\"{o}dinger equations and their solutions,  \newblock\href{https://doi.org/10.1017/S0334270000003891}{J. Austral. Math. Soc. Ser. B \textbf{25}, 16-43(1983).}
	\bibitem{Nail1} N. Akhmediev, J. M. Soto-Crespo, and A. Ankiewicz, Extreme waves that appear from nowhere: On the nature of rogue waves, \newblock\href{https://doi.org/10.1016/j.physleta.2009.04.023}{Phys. Lett. A \textbf{373}, 2137-2145 (2009).}
	\bibitem{exp0} D. R. Solli, C. Ropers, P. Koonath, and B. Jalali, Optical rogue waves, \newblock\href{https://doi.org/10.1038/nature06402}{Nature \textbf{450}, 1054 (2007).}
	\bibitem{exp1}B. Kibler, J. Fatome, C. Finot, G. Millot, F. Dias, G. Genty, N. Akhmediev, and J. M. Dudley, The Peregrine soliton in nonlinear fibre optics, \newblock\href{https://doi.org/10.1038/nphys1740}{Nat. Phys. \textbf{6}, 790 (2010).}
	\bibitem{exp2}A. Chabchoub, N. P. Hoffmann, and N. Akhmediev, Rogue Wave Observation in a Water Wave Tank, \newblock\href{https://doi.org/10.1103/PhysRevLett.106.204502}{Phys. Rev. Lett. \textbf{106}, 204502 (2011).}
	\bibitem{exp3}H. Bailung, S. K. Sharma, and Y. Nakamura, Observation of Peregrine Solitons in a Multicomponent Plasma with Negative Ions, \newblock\href{https://doi.org/10.1103/PhysRevLett.107.255005}{Phys. Rev. Lett. \textbf{107}, 255005 (2011).}
	\bibitem{Rev} J. M. Dudley, G. Genty, A. Mussot, A. Chabchoub, and F. Dias, Rogue waves and analogies in optics and oceanography, \newblock\href{https://doi.org/10.1038/s42254-019-0100-0}{Nat. Rev. Phys. \textbf{1}, 675-689 (2019).}
	\bibitem{ana3} Y. V. Bludov, V. V. Konotop, and N. Akhmediev, Matter rogue waves, \newblock\href{https://doi.org/10.1103/PhysRevA.80.033610}{Phys. Rev. A \textbf{80}, 033610 (2009).}
	\bibitem{ana4} A. Romero-Ros, G. C. Katsimiga, S. I. Mistakidis, B. Prinari, G. Biondini, P. Schmelcher, and P. G. Kevrekidis, Theoretical and numerical evidence for the potential realization of the Peregrine soliton in repulsive two-component Bose-Einstein condensates, \newblock\href{https://doi.org/10.1103/PhysRevA.105.053306}{Phys. Rev. A \textbf{105}, 053306 (2022).}
	\bibitem{Zhenya} Z. Yan, Vector financial rogue waves, \newblock\href{https://doi.org/10.1016/j.physleta.2011.09.026}{Phys. Lett. A \textbf{375}, 4274-4279 (2011).}
	\bibitem{Nail2} N. Akhmediev, A. Ankiewicz, and J. M. Soto-Crespo, Rogue waves and rational solutions of the nonlinear Schr\"{o}dinger equation, \newblock\href{https://doi.org/10.1103/PhysRevE.80.026601}{Phys. Rev. E \textbf{80}, 026601 (2009).}
	\bibitem{He1} L. H. Wang, K. Porsezian, and J. S. He, Breather and rogue wave solutions of a generalized nonlinear Schr\"{o}dinger equation, \newblock\href{https://doi.org/10.1103/PhysRevE.87.053202}{Phys. Rev. E \textbf{87}, 053202 (2013).}
	\bibitem{ling} B. Guo, L. Ling, and Q. P. Liu, Nonlinear Schr\"{o}dinger equation: Generalized Darboux transformation and rogue wave solutions, \newblock\href{https://doi.org/10.1103/PhysRevE.85.026607}{Phys. Rev. E \textbf{85}, 026607 (2012).}
	\bibitem{lingrw1} B.-L. Guo and L.-M. Ling, Rogue Wave, Breathers and Bright-Dark-Rogue Solutions for the Coupled Schr\"{o}dinger Equations, \newblock\href{http://iopscience.iop.org/0256-307X/28/11/110202}{Chin. Phys. Lett. \textbf{28}, 110202 (2011).}
	\bibitem{Baronio1} F. Baronio, A. Degasperis, M. Conforti, and S. Wabnitz, Solutions of the Vector Nonlinear Schr\"{o}dinger Equations: Evidence for Deterministic Rogue Waves, \newblock\href{https://doi.org/10.1103/PhysRevLett.109.044102}{Phys. Rev. Lett. \textbf{109}, 044102 (2012).}
	\bibitem{zhaorw1} L.-C. Zhao and  J. Liu, Localized nonlinear waves in a two-mode nonlinear fiber, \newblock\href{https://doi.org/10.1364/JOSAB.29.003119}{J. Opt. Soc. Am. B \textbf{29}, 3119 (2012).}
	\bibitem{zhaorw2} L.-C. Zhao and J. Liu, Rogue-wave solutions of a three-component coupled nonlinear Schr\"{o}dinger equation, \newblock\href{https://doi.org/10.1103/PhysRevE.87.013201}{Phys. Rev. E \textbf{87}, 013201 (2013).}
	\bibitem{zhaorw3} L.-C. Zhao, G.-G. Xin, and Z.-Y. Yang, Rogue-wave pattern transition induced by relative frequency, \newblock\href{https://doi.org/10.1103/PhysRevE.90.022918}{Phys. Rev. E \textbf{90}, 022918 (2014).}
	\bibitem{Liu1} C. Liu, S.-C. Chen, X. Yao, and N. Akhmediev, Non-degenerate multi-rogue waves and easy ways of their excitation, \newblock\href{https://doi.org/10.1016/j.physd.2022.133192}{Physica D \textbf{433}, 133192  (2022).}
	\bibitem{He2} J. S. He, H. R. Zhang, L. H. Wang, K. Porsezian, and  A. S. Fokas, Generating mechanism for higher-order rogue waves, \newblock\href{https://doi.org/10.1103/PhysRevE.87.052914}{Phys. Rev. E \textbf{87}, 052914 (2013).}
	\bibitem{ling2} L. Ling, B. Guo, and L.-C. Zhao, High-order rogue waves in vector nonlinear Schr\"{o}dinger equations, \newblock\href{https://doi.org/10.1103/PhysRevE.89.041201}{Phys. Rev. E \textbf{89}, 041201(R) (2014).}
	\bibitem{Liu2} C. Liu, Z.-Y. Yang, L.-C. Zhao, G.-G. Xin, and W.-L. Yang, Optical rogue waves generated on Gaussian background beam, \newblock\href{https://doi.org/10.1364/OL.39.001057}{Opt. Lett. \textbf{39}, 1057 (2014).}
	\bibitem{ling3} L.-C. Zhao, B. Guo, and L. Ling, High-order rogue wave solutions for the coupled nonlinear Schr\"{o}dinger equations-II, \newblock\href{https://doi.org/10.1063/1.4947113}{J. Math. Phys. \textbf{57}, 043508 (2016).}
	\bibitem{Gao} P. Gao, L.-C. Zhao, Z.-Y. Yang, X.-H. Li, and W.-L. Yang, High-order rogue waves excited from multi-Gaussian perturbations on a continuous wave, \newblock\href{https://doi.org/10.1364/OL.389012}{Opt. Lett. \textbf{45}, 2399 (2020).}
	\bibitem{MI1} C. Kharif and E. Pelinovsky, Physical mechanisms of the rogue wave phenomenon, \newblock\href{https://doi.org/10.1016/j.euromechflu.2003.09.002}{Eur. J. Mech. B \textbf{22}, 603-634 (2003).}
	\bibitem{MI} M. Onorato, S. Residori, U. Bortolozzo, A. Montina, and F. T. Arecchi, Rogue waves and their generating mechanisms in different physical contexts, \newblock\href{http://dx.doi.org/10.1016/j.physrep.2013.03.001}{Phys. Rep. \textbf{528}, 47-89 (2013).}
	\bibitem{MI2}J. M. Dudley, F. Dias, M. Erkintalo, and G. Genty, Instabilities, breathers and rogue waves in optics, \newblock\href{https://doi.org/10.1038/nphoton.2014.220}{Nature Photon \textbf{8}, 755-764 (2014).}
	\bibitem{zhaoMI} L.-C. zhao and L. Ling, Quantitative relations between modulational instability and several well-known nonlinear excitations, \newblock\href{http://dx.doi.org/10.1364/JOSAB.33.000850}{J. Opt. Soc. Am. B \textbf{33}, 850-856 (2016).}
	\bibitem{lingMI} L. Ling, L.-C. Zhao, Z.-Y. Yang, and B. Guo, Generation mechanisms of fundamental rogue wave spatial-temporal structure, \newblock\href{https://doi.org/10.1103/PhysRevE.96.022211}{Phys. Rev. E \textbf{96}, 022211 (2017).}
	\bibitem{defocusing1} F. Baronio, M. Conforti, A. Degasperis, S. Lombardo,  M. Onorato, and S. Wabnitz, Vector Rogue Waves and Baseband Modulation Instability in the Defocusing Regime, \newblock\href{https://doi.org/10.1103/PhysRevLett.113.034101}{Phys. Rev. Lett. \textbf{113}, 034101 (2014).}
	\bibitem{defocusing2} S. Chen, J. M. Soto-Crespo, and P. Grelu, Dark three-sister rogue waves in normally dispersive optical fibers with random birefringence, \newblock\href{https://doi.org/10.1364/OE.22.027632}{Opt. Express \textbf{22}, 27632 (2014).}
	\bibitem{chen1} S. Chen, F. Baronio, J. M. Soto-Crespo, P. Grelu, and D. Mihalache, Versatile rogue waves in scalar, vector, and multidimensional nonlinear systems, \newblock\href{https://doi.org/10.1088/1751-8121/aa8f00}{J. Phys. A: Math. Theor. \textbf{50}, 463001 (2017).}
	\bibitem{defocusing3} G. Zhang, Z. Yan, X.-Y. Wen, and Y. Chen, Interactions of localized wave structures and dynamics in the defocusing coupled nonlinear Schr\"{o}dinger equations, \newblock\href{https://doi.org/10.1103/PhysRevE.95.042201}{Phys. Rev. E \textbf{95}, 042201 (2017).}
	\bibitem{defocusing6} S. Chen, Y. Ye, J. M. Soto-Crespo, P. Grelu, and F. Baronio, Peregrine Solitons Beyond the Threefold Limit and Their Two-Soliton Interactions, \newblock\href{https://doi.org/10.1103/PhysRevLett.121.104101}{Phys. Rev. Lett. \textbf{121}, 104101 (2018).}
	\bibitem{defocusing5} B. Frisquet, B. Kibler, P. Morin, F. Baronio, M. Conforti, G. Millot, and S. Wabnitz, Optical Dark Rogue Wave, \newblock\href{https://doi.org/10.1038/srep20785}{Sci. Rep. \textbf{6}, 20785 (2016).}
	\bibitem{defocusing4} F. Baronio, B. Frisquet, S. Chen, G. Millot, S. Wabnitz, and B. Kibler, Observation of a group of dark rogue waves in a telecommunication optical fiber, \newblock\href{https://doi.org/10.1103/PhysRevA.97.013852}{Phys. Rev. A \textbf{97}, 013852 (2018).}
\bibitem{MI3} F. Baronio, S. Chen, P. Grelu, S. Wabnitz, and M. Conforti, Baseband modulation instability as the origin of rogue waves, \newblock\href{https://doi.org/10.1103/PhysRevA.91.033804}{Phys. Rev. A \textbf{91}, 033804 (2015).}
\bibitem{zhaoEPL} L.-C. Zhao, L. Duan, P. Gao, and Z.-Y. Yang, Vector rogue waves on a double-plane wave background, \newblock\href{https://doi.org/10.1209/0295-5075/125/40003}{EPL \textbf{125}, 40003 (2019).}	
\bibitem{Biondini1} G. Biondini and D. Kraus, Inverse scattering transform for the defocusing Manakov system with nonzero boundary conditions, \newblock\href{https://doi.org/10.1137/130943479}{SIAM J. Math. Anal. \textbf{47}, 706-757 (2015).}
\bibitem{lingdark} L. Ling, L.-C. Zhao, and B. Guo, Darboux transformation and multi-dark soliton for N-component nonlinear Schr\"{o}dinger equations, \newblock\href{http://iopscience.iop.org/0951-7715/28/9/3243}{Nonlinearity 28, 3243 (2015).}
\bibitem{Biondini2} G. Biondini, E. Fagerstrom, and B. Prinari, Inverse scattering transform for the defocusing nonlinear Schr\"{o}dinger equation with fully asymmetric non-zero boundary conditions, \newblock\href{https://doi.org/10.1016/j.physd.2016.04.003}{Physica D \textbf{333}, 117-136 (2016)}
\bibitem{PMI} B. Frisquet, B. Kibler, J. Fatome, P. Morin, F. Baronio, M. Conforti, G. Millot, and S. Wabnitz, Polarization modulation instability in a Manakov fiber system, \newblock\href{https://doi.org/10.1103/PhysRevA.92.053854}{Phys. Rev. A \textbf{92}, 053854 (2015).}	
	\bibitem{MIchen} S. Chen, L. Bu, C. Pan, C. Hou, F. Baronio, P. Grelu, and N. Akhmediev, Modulation instability-rogue wave correspondence hidden in integrable systems, \newblock\href{https://doi.org/10.1038/s42005-022-01076-x}{Commun. Phys. \textbf{5}, 297 (2022).}
\end{thebibliography}
\end{document}